\documentclass[preprint,proceedings]{rmaa}

\font\tenbg=cmmib10 at 10pt

\def \rvecphi{{\hbox{\tenbg\char'036}}}

\usepackage{paralist}
\usepackage{psfrag,color}

\SetYear{2005} \SetConfTitle{Triggering of Relativistic Jets}

\title{Launching of Poynting Jets from Accretion Disks}

\author{
   R. V. E. Lovelace,\altaffilmark{1}
M. M. Romanova,\altaffilmark{1}
 }

\altaffiltext{1}{Cornell University, Ithaca, NY USA.}

\shortauthor{Lovelace and Romanova}
\shorttitle{RevMexAA(SC) Poynting Jets}

\abstract{The jets observed to emanate from many
compact accreting objects may arise from the
twisting of the magnetic field threading a
differentially rotating accretion disk which
acts to magnetically extract angular
momentum and energy from the disk.
Two main regimes have been discussed, hydromagnetic
outflows, which have a significant mass flux
and have energy and angular momentum carried
by both matter and electromagnetic field and,
Poynting outflows, where the mass flux is
negligible and energy and angular momentum
are carried predominantly by the
electromagnetic field.
    We describe recent
theoretical work on the formation of
relativistic Poynting jets from magnetized accretion disks
and new relativistic, fully-electromagnetic,
particle-in-cell simulations  of the formation
of  jets from accretion disks.}

\addkeyword{AGN/Quasars: Jets}
\addkeyword{Compact Objects: Magnetic
Fields}

\begin{document}
\maketitle

\section{General}
\label{sec:intro}

Powerful, highly-collimated, oppositely
directed jets are observed in active galaxies
and quasars,
 and in old compact stars in binaries - the
`microquasars'.
Different models have been put
forward to explain astrophysical jets
(Bisnovatyi-Kogan \& Lovelace 2001).
Recent observational and theoretical work favors
models where twisting of an ordered
magnetic field threading an accretion disk acts to
magnetically accelerate the jets.
   Two main regimes have been considered in theoretical
models, the hydromagnetic regime
where the energy and angular momentum is carried by
both the electromagnetic field and
the kinetic flux of matter, and the Poynting flux
regime where the energy and angular
momentum outflow from the disk is carried
predominantly by the electromagnetic field.
    In \S 3 we outline the theory of  Poynting jets.
  In \S 4 we present new results from axisymmetric, fully
electromagnetic,
relativistic-particle-in-cell (PIC)
simulations of the formation and propagation
of relativistic jets from a disk.

\begin{figure}[!t]
  \includegraphics[width=\columnwidth]{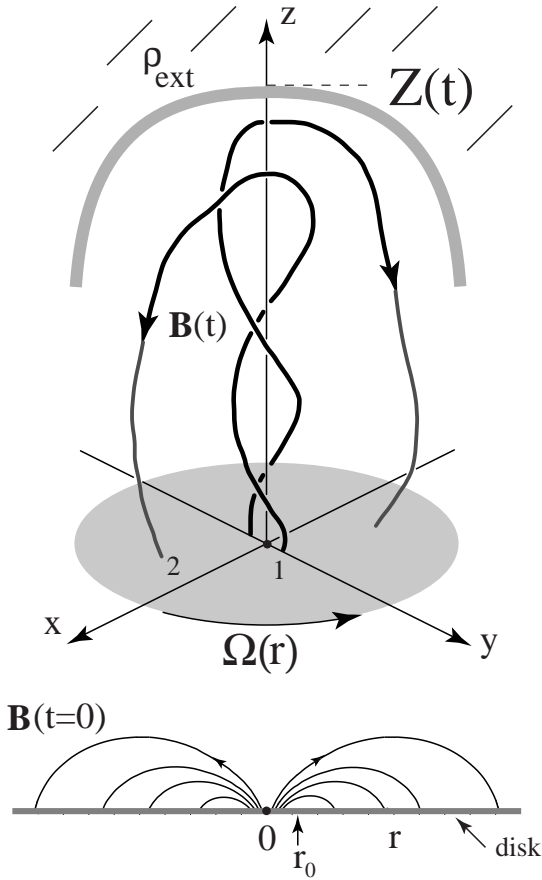}
  \caption{
Sketch of the magnetic field
configuration of a Poynting jet from Lovelace and
Romanova (2003). The bottom part of the figure
shows the initial dipole-like magnetic field
threading the disk which rotates at the angular
rate $\Omega(r)$. The top part of the figure shows
the jet at some time later when the head of the
jet is at a distance $Z(t)$. At the head of the jet
there is force balance between electromagnetic
stress of the jet and the ram pressure of the
ambient medium of density $\rho_{ext}$.
}
  \label{fig:simple}
\end{figure}

\section{Theory of Poynting Jets}

   The powerful jets observed from active
galaxies and quasars are probably not
hydromagnetic outflows but rather Poynting flux
dominated jets.
   The motion of these jets measured
by very long baseline interferometry correspond
to bulk Lorentz factors of $\Gamma = {\cal O}(10)$
which is much larger than the Lorentz
factor of the Keplerian disk velocity
predicted for hydromagnetic outflows.
  Furthermore, the low Faraday rotation measures
observed for these jets at distances $<$ kpc from
the central object implies a very low plasma
densities. Similar arguments indicate that the
jets of microquasars are not hydromagnetic
outflows but rather Poynting jets. Poynting Jets
have been proposed to be the driving mechanism for
gamma ray burst sources (Katz 1997).
Theoretical studies have developed models for
Poynting jets from accretion disks (Lovelace,
Wang, \& Sulkanen 1987; Lynden-Bell  2003;
Romanova \& Lovelace 1997; Levinson 1998;
and
Lovelace {\it et al.} 2002; and Lovelace
\& Romanova 2003 ).
   Stationary Poynting flux
dominated outflows were found by Romanova {\it et al.}
(1998) and Ustyugova {\it et al.} (2000) in
axisymmetric MHD simulations of the opening of
magnetic loops threading a Keplerian disk.

        We first summarize the theory of
non-relativistic Poynting jets which is based on
the Grad-Shafranov equation.
  We show results of
non-relativistic, axisymmetric MHD simulations
which support this theory.
Later, we discuss the corresponding results
obtained by solving the relativistic
Grad-Shafranov equation.

        Consider the coronal magnetic
field - such as that shown in the lower part of
Figure 1 - of a differentially rotating Keplerian
accretion disk. That is, the disk is perfectly
conducting, high-density, and has a small
accretion speed ($\ll v_K$). Further, consider
``coronal'' or ``force-free" magnetic fields in the
non-relativistic limit.
  We use cylindrical $(r, \phi,
z)$ coordinates and consider axisymmetric field
configurations. Thus the magnetic field has the
form ${\bf B} = {\bf B}_p + B_\phi\hat{\rvecphi~}$,
with ${\bf B}_p =
B_r\hat{\bf r} + B_z\hat{\bf z}$.
Because ${\bf \nabla \cdot B} = 0$, ${\bf B}
 = {\bf \nabla \times A}$
with ${\bf A}$ the vector potential.
  Consequently, $B_r = -(1/r)\partial \Psi/\partial z$
and  $B_z = (1/r)\partial \Psi/\partial r$.
where $\Psi(r,z) \equiv rA_\phi(r,z)$.
   The $\Psi(r,z) =$ const
lines label the poloidal field lines; that is,
$({\bf B}_p \cdot {\bf \nabla})\Psi=0 =({\bf B}\cdot
{\bf \nabla}) \Psi$.
   Note that $2\pi \Psi(r, z)$ is
the magnetic flux through a horizontal, coaxial
circular disk of radius $r$.
  The magnetic field
threading the disk at $z=0$ is assumed to evolve
slowly so that it can be considered approximately
time-independent, $\Psi(r,z=0) = \Psi_0(r)$. However, the
magnetic field above the disk will in general be
time-dependent, $\Psi= \Psi(r, z, t)$, due to the
differential rotation of the disk.

        The non-relativistic equation of
plasma motion in the corona of an accretion disk
is $\rho d{\bf v}/dt = -{\bf \nabla }p +
\rho {\bf g} + {\bf J \times B}/c$, where $\bf v$
is the flow velocity, $p$ is the pressure,
and $\bf g$ is the
gravitational acceleration.
  The equation for the
$\bf B$ field is ${\bf \nabla \times B}
= 4\pi {\bf J}/c$, because the
displacement current is negligible in the
non-relativistic limit. In the coronal or
force-free plasma limit, the magnetic energy
density ${\bf B}^2/8\pi $ is much larger than the kinetic
or thermal energy densities; that is, we have
sub-Alfv\'enic flow speeds
${\bf v}^2 \ll v_A^2 ={\bf B}^2/(4\pi\rho)$,
where $v_A$ is the Alfv\'en velocity.
The force equation then simplifies to $0 \approx
{\bf J \times B}$ so that
${\bf J}= \lambda{\bf B}$ (Gold \& Hoyle 1960).
   Because ${\bf \nabla}\cdot{\bf J}=0$,
$({\bf B}\cdot{\bf \nabla}) \lambda=0$, and
consequently $\lambda=\lambda(\Psi)$, as well-known.
   Thus Amp\`ere's law becomes
${\bf \nabla \times B}=4\pi \lambda(\Psi){\bf B}/c$.
The $r$ and $z$ components
of this equation imply $r B_\phi =
H(\Psi)$, and $dH(\Psi)/d\Psi=4\pi\lambda(\Psi)/c$,
where $H(\Psi)$ is
another function of $\Psi$.
  Thus, $H(\Psi) =$ const are
lines of constant poloidal current density;
${\bf J}_p =(c/4\pi) (dH/d\Psi){\bf B}_p$
so that $({\bf J}_p\cdot{\bf \nabla})H=0.$
The toroidal component of
Amp\`ere s law gives the non-relativistic
Grad-Shafranov  equation for $\Psi$,
$$
\Delta^*\Psi = -H(\Psi) {d H(\Psi) \over d\Psi}~.
\eqno(1)
$$
   Here, $\Delta^* \equiv \partial^2/\partial r^2
-(1/r)(\partial/\partial r)+\partial^2/\partial z^2$
is the adjoint Laplacian
operator.
    Note that $\Delta^*\Psi=r(\nabla^2-1/r^2)A_\phi$ and
that $H(dH/d\Psi) = 4\pi r J_\phi/c$.
   From Amp\`ere's law, $\oint d{\bf l}\cdot {\bf B}=
(4\pi/c)\int d{\bf S}\cdot {\bf J}$,
so that $rB_\phi(r, z) = H(\Psi)$
 is $(2/c)$ times the current flowing through a
circular area of radius $r$
(with normal $\hat{\bf z}$) labeled
by $\Psi(r,z)=$ const.
  Equivalently, $-H[\Psi(r, 0)]$ is
$(2/c)\times$ the current flowing into the area of the
disk with radii $\leq r$.
  For all cases studied here, $-H(\Psi)$ has a
maximum so that the total current flowing into the
disk for $r \leq  r_m$ is $I=(2/c)(-H)_{max}$, where $r_m$
is such that $-H[\Psi(r_m,0)] =(-H)_{max}$ so that $r_m$ is less
than the radius of the O-point, $r_0$. The same total current
$I_{tot}$ flows out of the region of the disk $r = r_m$ to $r_0$.

    The function $H(\Psi)$ must be
determined before the Grad-Shafranov
equation can be solved.
   $H(\Psi)$ is determined by the
differential rotation of the disk:
The azimuthal twist of a
given field line going from an inner footpoint at $r_1$ to an
outer footpoint at $r_2$ is fixed by the differential rotation
of the disk.
  The field line slippage speed through the disk
due to the disk's finite magnetic diffusivity is estimated
to be negligible compared with the Keplerian velocity $v_K$.
  For a given field line we have $r d\phi /B_\phi = ds_p/B_p$, where
$ds_p =\sqrt{dr^2 +dz^2}$ is the poloidal arc length along the
field line, and $B_p = \sqrt{B_r^2 + B_z^2}$.
The total twist of a field line loop is
$$
\Delta \phi(\Psi) = -\int_1^2 ds_p~ {B_\phi \over r B_p}
=-H(\Psi)\int_1^2 {d s_p \over r^2 B_p}~,
\eqno(2)
$$
with the sign included to give $\Delta \phi > 0$.
For a Keplerian disk around an object of mass $M$, the angular
rotation rate is $\Omega_K = \sqrt{GM/r^3}$
 so that the field line
twist after a time $t$ is $\Delta \phi(\Psi)
 = \Omega_0  t \big[(r_0/r_1 )^{3/2} -
(r_0 / r_2 )^{3/2}\big] = (\Omega_0t) F(\Psi/\Psi_0)$,
where $r_0$ is the radius of the
O-point, $\Omega_0 =\sqrt{GM/r_0^3}$, and $F$ is a dimensionless
function.

\begin{figure}[!t]
  \includegraphics[width=\columnwidth]{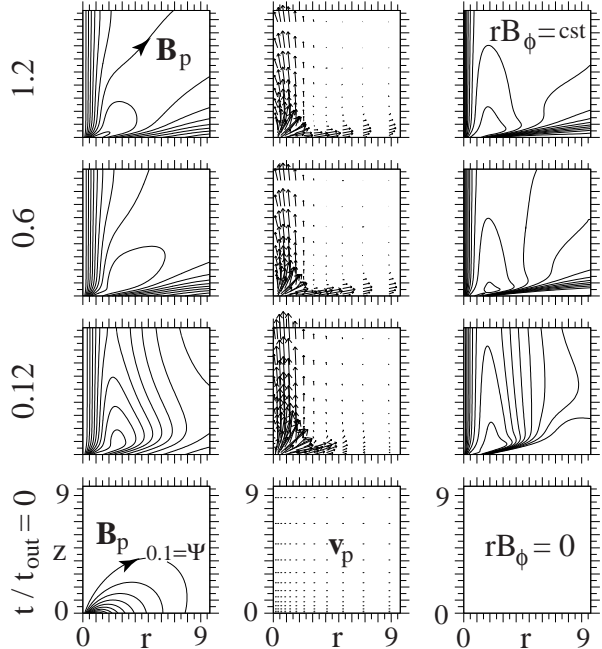}
  \caption{
Non-relativistic time evolution
of dipole-like field threading
the disk from the initial
configuration $t = 0$ (bottom panels) to
the final quasi-stationary state
at $t = 1.2t_{out}$, where $t_{out}$ is the
rotation period of the disk at the
outer radius $R_{out}$ of the simulation
region from Ustyugova {\it et al.} (2000).
   The left hand panels show the
poloidal field lines which are the
same as the $\Psi(r,z) =$const lines; $\Psi$ is
normalized by $\Psi_{max}$, and the spacing
between lines is $0.1$.
  The middle
panels show the poloidal velocity
vectors ${\bf v}_p$. The right-hand panels show
the constant lines of $-rB_\phi(r,z) > 0$
in units of $\Psi_{max}/r_0$, and the spacing
between lines is $0.1$. For this
calculation a $100 \times 100$ inhomogeneous
grid was used with $\Delta r_j$ and $\Delta z_k$
growing with
    distance $r$ and $z$ geometrically as
$\Delta r_j = \Delta r_1 q^j$
and $\Delta z_k = \Delta z_1 q^k$,
with $q = 1.03$ and $\Delta r_1 = \Delta z_1
= 0.05r_0$.
}
 \label{fig:simple}
\end{figure}

        Equations (1) and (2) have been solved numerically by Li
{\it et al.} (2001) and Lovelace {\it et al.} (2002) for an initial
poloidal magnetic field as shown in the lower part of
Figure 1.
   As the ``twist,'' as measured
by $\Omega_0 t$, increases,
a high twist field configuration appears with a different
topology.
    A ``plasmoid'' consisting of toroidal flux
detaches from the disk and propagates outward. The plasmoid
is bounded by a poloidal field line which has an X-point
above the O-point on the disk.
  The occurrence of the
X-point requires that there be at least a small amount of
dissipation in the evolution from the poloidal dipole field
and the Poynting jet configuration.
  The high-twist
configuration consists of a region near the axis which is
magnetically collimated by the toroidal $B_\phi$ field and a
region far from the axis which is anti-collimated in the
sense that it is pushed away from the axis.
   The field lines
returning to the disk at $r > r_0$ are anti-collimated by the
pressure of the toroidal magnetic field. The poloidal field
fills only a small fraction of the coronal space.

        Figure 2 shows results of non-relativistic axisymmetric
simulations of the formation of a Poynting jet by Ustyugova
{\it et al.} (2000).
   The flow near the $z-$axis is the Poynting jet
and its physical properties agree with Grad-Shafranov
solutions.

In the case of relativistic Poynting jets we hypothesize
that the magnetic field
configuration is similar to that in the non-relativistic limit
(Ustyugova {\it et al.} 2000; Lovelace
    {\it et al.} 2002).
    Thus, most of the twist $\Delta \phi$ of a field line
of the relativistic Poynting jet
    occurs along the jet from $z = 0$ to $Z(t)$ as sketched in
Figure 3, where $Z(t)$ is the
    axial location of the ``head'' of the jet.
     Along most of the
distance $z = 0$ to $Z$, the radius
    of the jet is a constant and
$\Psi = \Psi(r)$ for $Z >> r_0$.
    Note that the function $\Psi(r)$
is different from $\Psi(r,0)$ which
is the flux function profile
on the disk surface.
     Hence $r^2 d\phi/dz =rB_\phi(r,z)/B_z(r,z)$.
 We take for simplicity  $V_z =
dZ/dt =$ const.
     We determine $V_z$ subsequently.
  In this case  $H (\Psi) =[r^2\Omega(\Psi)/V_z]B_z$
can be written as a function of $\Psi$ and
$d\Psi/dr$.  With $H$ known, the relativistic
Grad-Shafranov equation,
$$
\left[1\!-\left({r\Omega \over c}\right)^2\right]\!
\Delta^*\Psi -{{\bf \nabla}\Psi\over 2r^2}\cdot
{\bf \nabla}\left({r^4\Omega^2 \over c^2}\right)\!=
-H(\Psi){d H(\Psi) \over d \Psi},
\eqno(3)
$$
can be solved (Lovelace \& Romanova 2003).

\begin{figure}[!t]
  \includegraphics[width=\columnwidth]{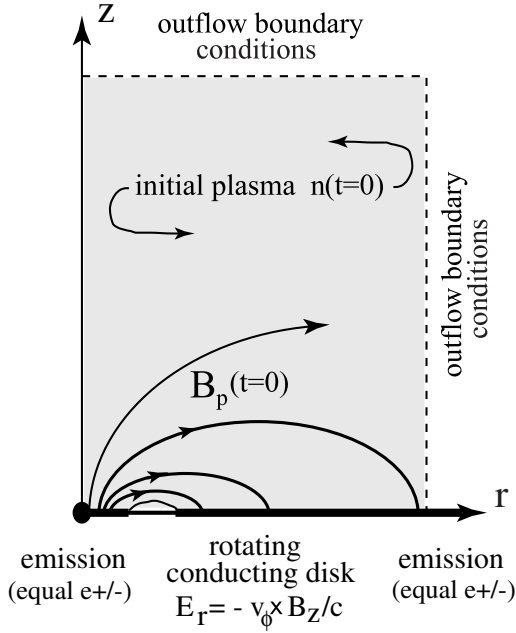}
  \caption{
Sketch of the initial conditions
for the relativistic particle-in-cell simulations of
jet formation from an accretion disk.
}
 \label{fig:simple}
\end{figure}

      The quantity not determined by
equation (3) is the velocity $V_z$, or Lorentz factor
$\Gamma =1/(1-V_z^2/c^2)^{1/2}$.
 This is determined by taking into account the balance of
axial forces at the head of the jet: the electromagnetic
pressure within the jet is balanced against the
dynamic pressure of the external medium
which is assumed uniform with density $\rho_{ext}$.
This gives $(\Gamma^2-1)^3=B_0^2/
(8\pi {\cal R}^2\rho_{ext} c^2)$, or for $\Gamma \gg 1$,
$$
\Gamma \approx 8\left({10 \over {\cal R}}\right)^{1/3}
\left({B_0 \over 10^3 {\rm G}}\right)^{1/3}
\left({1/{\rm cm}^3 \over n_{ext}}\right)^{1/6}~,
\eqno(4)
$$
where ${\cal R} = r_0/r_g \gg 1$, with $r_0$ the
O-point of the magnetic field, $r_g \equiv  GM/c^2$, and
$B_0$ the magnetic field strength at the
center of the disk. This value of $\Gamma$ is of the
order of the Lorentz factors of the expansion of
parsec-scale extragalactic radio jets observed
with very-long-baseline-interferometry
(see, e.g., Zensus {\it et al.} 1998).
This interpretation assumes that the
radiating electrons (and/or positrons)
are accelerated to high Lorentz
factors ($\gamma \sim 10^3$) at the
jet front and move with a
bulk Lorentz factor $\Gamma$ relative to the
observer.
   The luminosity of the $+z$ Poynting
jet is $\dot{E}_j =c\int_0^{r_2}
rdrE_rB_\phi/2= cB_0^2{\cal R}^{3/2}r_g^2/3\sim
2.1 \times 10^{46} (B_0/ 10^3{\rm G})^2
({\cal R}/10)^{3/2}(M/10^9M_\odot)^2$ erg/s,
where $M$ is the mass of
the black hole.

        For long time-scales, the Poynting jet
is of course time-dependent due to the
angular momentum it extracts from
the inner disk ($r < r_0$) which in turn
causes $r_0$ to decrease with time
(Lovelace {\it et al.} 2002).
  This loss of angular momentum
leads to a ``global
magnetic instability'' and collapse
of the inner disk (Lovelace {\it et al.}
1994, 1997, 2002) and a corresponding
outburst of energy in the jets from
the two sides of the disk.
  Such outbursts may explain the flares of
active galactic nuclei blazar sources
(Romanova \& Lovelace 1997; Levinson
1998) and the one-time outbursts of
gamma ray burst sources (Katz 1997).

\section{Relativistic Particle-in-Cell Simulations
of Jets}
\medskip

    We performed relativistic, fully
electromagnetic, particle-in-cell
simulations of the formation of relativistic
 jets from  an accretion
disk initially threaded by a dipole-like magnetic field.
   This was done using the code XOOPIC developed
by Verboncoeur, Langdon, and Gladd (1995).
  Earlier, Gisler, Lovelace, and Norman (1989)
studied jet formation for a monopole type field using
the relativistic PIC code ISIS.
   The geometry of the initial configuration is
shown in Figure 3.
    The computational region is a cylindrical ``can,''
$r=0 -R_m$ and $z=0 - Z_m$, with outflow boundary conditions
on the outer boundaries, and the potential and particle
emission specified on the disk surface $r=0-R_m$, $z=0$.
  Equal fluxes of electrons and positrons are emitted
so that the net emission  is
effectively  space-charge-limited.
  About $10^5$ particles were used in the simulations
reported here.
   The behavior of the lower half-space ($z<0$) is expected to
be a mirror image of the upper half-space.

   Figure 4 shows the formation
of a relativistic jet.
   The gray scale indicates the logarithm
of the density of electrons or positrons with $20$ levels
between the lightest ($10^{12}$)
and darkest ($4\times 10^{15}$/m$^3$).
   The lines are poloidal
magnetic field lines ${\bf B}_p$.  The total,
three-dimensional magnetic field
is  shown in Figure 5.
  The computational region has
$(R_m,Z_m)=(50,100)$ m, the initial ${\bf B}-$field is
dipole-like  with
$B_z(0,0)\equiv B_0=28.3$ G and an O-point at
$(r,z)=(10, 0)$ m, and the electric
 potential at the center of the disk
is $\Phi_0=-10^7$ V relative to the outer region of the
disk.
   Initially, the computational region was filled
with a distribution of equal densities of electrons
and positrons with $n_\pm(0,0) =3\times 10^{13}$/m$^3$.
  Electrons and positrons
are emitted with equal currents $I_{\pm} = 3\times 10^5$ A
from both the inner and the outer
portions of the disk as indicated in Figure 3 with an
axial speed much less than $c$.
   For a Keplerian disk with $r_0 \gg r_g$, the scalings are $\Phi_0
\sim  B_0 (r_0 r_g)^{1/2}$,
$I\sim cB_0 r_0$ and the jet power is $\sim cB_0^2
r_0^{3/2}r_g^{1/2}$.
    The calculations
were done on a $64\times 128$ grid stretched in
both the $r$ and $z$ directions so as to give
much higher spatial resolution at small $r$ and small $z$.
  These simulations show the formation of a quasi-stationary,
collimated current-carrying jet.
    The Poynting flux
power of the jet is $\dot{E}_j\approx 7\times 10^{11}$ W and the
particle kinetic energy power is $\approx 4.7 \times 10^{10}$ W.
  The charge density
of the electron flow is partially neutralized by
the positron flow.
    Simulations are planned with the positrons
replaced by ions.

\begin{figure}[!t]
  \includegraphics[width=\columnwidth]{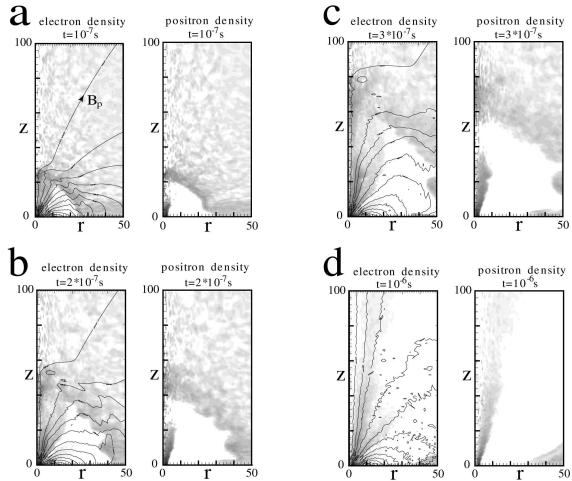}
  \caption{
Relativistic particle-in-cell simulations of
the formation of a jet from a rotating disk.
(a) -(c) give snapshots at times $(1,2,3)\times10^{-7}$ s,
and (d) is at $t=10^{-6}$ s.
}
\label{fig:simple}
\end{figure}

\begin{figure}[!t]
  \includegraphics[width=\columnwidth]{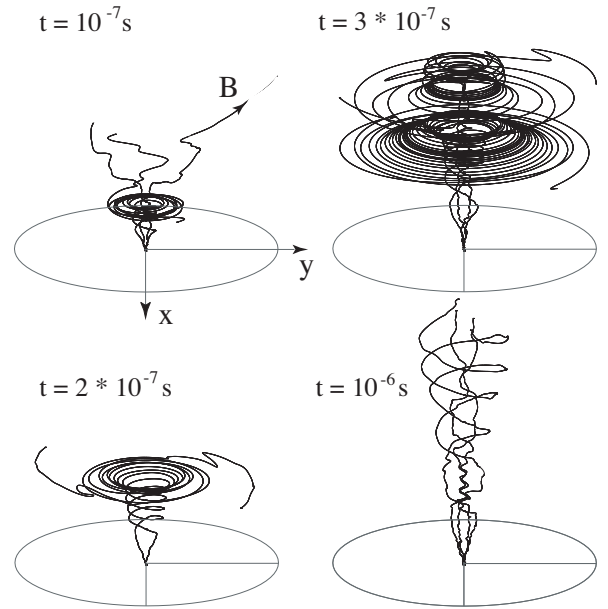}
  \caption{
Three dimensional magnetic
field lines originating from the disk
at $r=1,~ 2$ m  for the same case as Figure 4.
}
  \label{fig:simple}
\end{figure}

\acknowledgements

   We thank the meeting organizers for the stimulating and
very well organized meeting.
  This work was supported in part by DOE cooperative
agreement DE-FC03 02NA00057.

\end{document}